\documentclass[pra,showpacs,aps,amssymb]{revtex4}
\usepackage{amsmath}

\def\be{\begin{equation}}
\def\te{\end{equation}}
\def\ee{\end{equation}}
\def\ba{\begin{eqnarray}}
\def\bea{\begin{eqnarray}}
\def\nn{\nonumber\\}
\def\tea{\end{eqnarray}}
\def\ea{\end{eqnarray}}
\def\eea{\end{eqnarray}}

\begin{document}

\title{Kadanoff-Baym equations for near-Kolmogorov turbulence}
\author{Esteban Calzetta\footnote{e-mail: calzetta@df.uba.ar} }
\affiliation{Depto. de F\'{\i}sica, F. C. E. y N. - UBA and CONICET
\\
 pab. I, Cdad. Universitaria
\\ 
(1428) Buenos Aires, Argentina}
\date{\today }

\begin{abstract}
We use the 2 particle irreducible Schwinger-Keldysh effective action to set up consistent equations for the velocity and pressure correlations of a turbulent flow. We use these equations to derive the Kadanoff-Baym equations describing the relaxation to Kolmogorov turbulence in the absence of mean velocities.
\end{abstract}

\maketitle

\section{Introduction}
Since the second half of last century field theory techniques have been applied to an ever widening set of problems. Turbulence theory has featured prominently in these efforts. In fact, this field has been so active that it would be impossible to give here a comprehensive set of references; we refer the reader instead to the monographies \cite{MonYag71} and \cite{MCCO94}, and the review articles \cite{LvoPro95,BLPP98}. In spite of all this effort, and without disregarding the fact that there are diverging opinions about what exactly has been accomplished \cite{FRIS95,Tsin04}, it can be argued that the array of tools from field theory which are relevant to turbulence has not been exhausted yet \cite{CalHu08}.

As a concrete example, let us consider Gioia and Chakravorty's ansatz for the friction factor in turbulent pipe flow \cite{GioCha06,Cal09}. Gioia and Chakravorty argue that a global feature of the flow (the friction it produces on the walls containing the flow) is directly determined by the spectrum of turbulent energy at small scales within the flow. What is at stake is how ``hard'' (short wavelength) excitations of the system manage to affect the ``soft'' (long wavelength) scales (see also \cite{LLPP05}). This is a typical problem in the theory of non abelian plasmas, such as the one produced after a relativistic heavy ion collision \cite{BlaIan02,LitMan02}. In the high energy  context, this problem is usually handled by deriving Kadanoff-Baym kinetic equations for the correlations of the relevant fields \cite{KadBay62}. While not unknown \cite{EdwMcC72}, this kind of approach has not been extensively developed in turbulence theory, except in the theory of wave turbulence \cite{ZaLvFa92}.

The Kadanoff-Baym equations themselves are most efficiently derived as an approximation to a more detailed set of Schwinger-Dyson equations for the relevant correlations. The Schwinger-Dyson equations in turn may be derived from the variation of the two particle irreducible (2PI) Schwinger - Keldysh (or closed time-path, CTP)  effective action (EA) \cite{CalHu08}.

The different elements of this approach have been discussed in more or less detail in the turbulence literature: the Schwinger-Keldysh technique, or rather the closely related Martin-Siggia-Rose formalism \cite{MaSiRo73,HorLip79,ZanCal02}, the use of generating functionals (whose Legendre transform is the effective action) and diverse resummation schemes \cite{BelLvo87,MCCO94,LvoPro95,Kiy04}. However, they have been rarely applied together. Moreover, we are not interested in the 2PI approach as a way to resum a perturbative expansion for the correlations. Our goal is to elucidate what can be learnt about fully developed turbulence from the fact that the main equations \emph{can} be derived from a variational principle.

The subject of this paper is the formulation of the Schwinger-Dyson equations for the mean velocity and pressure, and the two-time, two points velocity-velocity, velocity-pressure and pressure-pressure correlations for incompressible flow as derived from the 2PI CTP EA \cite{CalHu08}. For simplicity we shall consider  flow in an unbounded domain, and we will be mostly concerned with the regime of very large Reynolds number $Re$. As an application we shall derive a transport equation describing the evolution of a nearly homogeneous flow (we define below what we mean by nearly homogeneous) towards a Kolmogorov-like cascade. These are highly formal issues, but they need to be disposed of before more physical questions, such as the effect of boundaries and/or shear on the flow may be considered.

This paper is organized as follows. In next Section we present the problem and discuss the notation to be used henceforth.

Section III is devoted to the subject of random Galilean invariance. This symmetry, which to the best of our knowledge has been introduced by Kraichnan \cite{Kra65}, is the most important structural feature of the model. See \cite{HorLip79,BelLvo87,Spe88,MCCO94,BLPP98,BerHoc07}. We shall show that the self-energies introduced in Section II, which represent all nonlinear effects on the dynamics of correlations, are naturally random Galilean invariant (RGI). This holds even when the Navier-Stokes equations themselves are not RGI, due do the random external forces. The fact that this important result is almost trivial in this formalism is probably the most compelling argument that could be offered in its favour.

In Section IV we discuss the general features of the correlation dynamics in the case where the mean velocity and pressure vanish. As an example, we describe the solution to the Schwinger-Dyson equations corresponding to Kolmogorov scaling.

Finally, in Section V we introduced nearly homogeneous flows, derive the local transport equation which describes their dynamics and use it to discuss the approach towards Kolmogorov scaling. This is probably the only truly new result in this paper. We present it as a simple application to better judge all the machinery introduced so far. We conclude with some brief final remarks.

We have added two appendices: Appendix A discusses an approximation made in Section VI, and Appendix B shows how the  K\'{a}rm\'{a}n-Howarth equation \cite{KarHow38,MonYag71,FRIS95} and the Kolmogorov $4/5$ Law \cite{FRIS95,Ras99} may be derived from the 2PI CTP EA formalism.  It seems a fair demand on a new formalism that the main successes of the old one should be reproduced without uncommon difficulty.

\section{Definitions and notations}
In this Section we define the model, introduce the 2PI CTP EA, show that only a subset of the solutions to the Schwinger-Dyson equations represent physical flows and, most importantly, settle on the notation to be used in what follows.

\subsection{Navier-Stokes equations for incompressible fluids}
We consider a flow defined on unbounded $d$-dimensional space. We shall work in eulerian coordinates throughout. At each point $\mathbf{x}$ and time $t$ the flow is defined by the components $\mathbf{U}^p\left[\mathbf{x},t\right]$ of the velocity and the pressure $P\left[\mathbf{x},t\right]$. 

The fluid is subject to an external random force $\mathbf{f}_p\left[\mathbf{x},t\right]$. It is one of the basic assumptions of turbulence theory that at high enough Reynolds number $Re$ the main features of the flow are robust with respect to the forcing. We therefore have considerable freedom regarding the choice of the stochastic process $\mathbf{f}_p\left[\mathbf{x},t\right]$. We shall assume it is Gaussian, with zero mean and self correlation

\be
\left\langle \mathbf{f}_p\left[\mathbf{x},t\right]\mathbf{f}_q\left[\mathbf{y},t_y\right]\right\rangle =c\mathbf{N}_{pq}\left[\mathbf{x},t;\mathbf{y},t_y\right]
\label{eq1}
\te
The constant $c=L^{d}V^{2}$, where $L$ and $V$ are macroscopic length and velocity scales, is introduced for later purposes; observe that $c$ has dimensions. 

There is an important class of forcings, those where

\be
\mathbf{N}_{pq}\left[\mathbf{x},t;\mathbf{y},t_y\right]=\mathbf{N}^{eq}_{rs}\left[\mathbf{x}-\mathbf{y}\right]\delta\left(t-t_y\right)
\label{eq2}
\te
These make the Navier-Stokes equations random Galilean invariant (RGI) (see next Section). We shall not assume the noise self-correlation \ref{eq1} has the form \ref{eq2} unless explicitly noted (this will happen in Section V).

The fluid is incompressible, meaning that

\be
\nabla_{p}\mathbf{U}^p=0
\label{eq3}
\te
The dynamics of the flow is described by the incompressible Navier-Stokes equations

\be
\left[\frac{\partial}{\partial t}-\nu_b\nabla^2\right]\mathbf{U}^p+\nabla_{q}\left[\mathbf{U}^p\mathbf{U}^q\right]+\nabla^pP=\mathbf{f}^p
\label{eq4}
\te
where $\nu_b$ is the kinematic viscosity. 

The natural notation we have used so far is too involved for the discussion to come. We shall compress it as follows. We first compress the space and time dependence into a single continuous index $a$

\be
P\left[\mathbf{x},t\right]\mapsto P^a
\label{eq5}
\te
For the velocity, we replace both the space-time dependence and component index $p$ into a single index $j$

\be
\mathbf{U}^p\left[\mathbf{x},t\right]\mapsto \mathbf{U}^j
\label{eq6}
\te
The equations of motion \ref{eq3} and \ref{eq4} take the form

\be
\mathbf{D}_j\left[\mathbf{U},P\right]=D_{jk}\mathbf{U}^k+L_{jb}P^b+g_{jkl}\mathbf{U}^k\mathbf{U}^l=\mathbf{f}_j
\label{NS}
\te

\be
L_{jb}\mathbf{U}^j=0
\label{incomp}
\te
Here $j=\left(\mathbf{x},t,p\right)$, $k=\left(\mathbf{y},t_y,q\right)$, $l=\left(\mathbf{z},t_z,r\right)$ and $b=\left(\mathbf{y},t_y\right)$. Further

\be
D_{jk}=\delta_{pq}\left[\frac{\partial}{\partial t}-\nu_b\nabla^2\right]\delta\left(t-t_y\right)\delta\left(\mathbf{x}-\mathbf{y}\right)
\label{eq7}
\te

\be
L_{jb}=\nabla_p \delta\left(t-t_y\right)\delta\left(\mathbf{x}-\mathbf{y}\right)
\label{eq8}
\te

\be
g_{jkl}=\frac12\left[\delta_{pq}\nabla_r+\delta_{pr}\nabla_q\right]\delta\left(t-t_y\right)\delta\left(\mathbf{x}-\mathbf{y}\right)\delta\left(t-t_z\right)\delta\left(\mathbf{x}-\mathbf{z}\right)
\label{eq9}
\te
All expressions are summed (integrated) over repeated discrete (continuous) indexes.

\subsection{The CTP generating functional}

Following the time-honored procedure we decompose all fields in mean field and fluctuations $\mathbf{U}^j=\mathbf{\bar{u}}^j+\mathbf{u}^j$ and $P^a=\bar{p}^a+p^a$. We wish to find self-consistent equations for the mean fields and the two-time correlators

\be
G^{jk}=\left\langle \mathbf{u}^j\mathbf{u}^k\right\rangle
\te

\be
H^{ja}=\left\langle \mathbf{u}^jp^a\right\rangle
\te

\be
I^{ab}=\left\langle p^ap^b\right\rangle
\te
The idea is to write both men fields and correlations as derivatives of a generating functional

\be
\mathbf{\bar{u}}^j=\frac{\delta W}{\delta J^+_j}
\te
etc., where

\be
e^{iW/c}=\left\langle e^{iS_J/c}\right\rangle
\te
The brackets denote an ensemble average, and $S_J$ is the source action

\be
S_J=J^{+}_j\mathbf{U}^j+J^{+}_aP^a+\frac12K^{++}_{jk}\mathbf{U}^j\mathbf{U}^k+K^{++}_{ja}\mathbf{U}^jP^a+\frac12K^{++}_{ab}P^aP^b
\te
By adding the constant $c$ from \ref{eq1}, the new source $J^{+}_j$ has the same dimensions as the physical random source $\mathbf{f}_j$. More explicitly

\be
e^{iW/c}=\int\:D\mathbf{f}D\mathbf{U}DP\;\mathcal{P}\left[\mathbf{f} \right]\delta\left(\mathbf{D}_j\left[\mathbf{U},P\right]-\mathbf{f}_j\right)\delta\left(L^{\dagger}_{aj}\mathbf{U}^j\right)\:e^{iS_J/c}
\label{fungen}
\te
where $\mathcal{P}$ is the probability density functional for the random source $\mathbf{f}$. We shall call the fields in \ref{fungen} the physical fields and denote them by a $+$ superindex. We introduce mirror fields $\mathbf{U}^{j-}$ and $P^{a-}$ to exponentiate the delta functions. The mirror fields are defined to have the same dimensions as the corresponding physical field. By symmetry, we add sources to $S_J$ coupled to mirror fields, products of two mirror fields, and products of one physical and one mirror field.

The result of this activity is a theory of field doublets $\mathbf{U}^J=\left(\mathbf{U}^{j+},\mathbf{U}^{j-}\right)$ and $P^A=\left(P^{a+},P^{a-}\right)$. Observe that the new index $J$ combines the old $j$ of the velocity field and the new index $\alpha =\pm$ distinguishing between physical and mirror fields, $J=\left(\mathbf{x},t,r,\alpha\right)$. Because of Schwinger-Keldysh tradition, we shall call $\alpha$ the ``branch'' index \cite{CalHu08}. Similarly the new index $A$ combines the space-time and branch labels $A=\left(\mathbf{x},t,\alpha\right)$. 

Integrating over the random sources we get

\be
e^{iW/c}=\int\:D\mathbf{U}DP\;e^{i\left[S+S_J\right]/c}
\label{genfun}
\te
where

\be
S=\mathbf{U}^{j-}\left[D_{jk}\mathbf{U}^{k+}+g_{jkl}\mathbf{U}^{k+}\mathbf{U}^{l+}\right]+\mathbf{U}^{j-}L_{ja}P^{a+}+\mathbf{U}^{j+}L_{ja}P^{a-}+\frac i2\mathbf{U}^{j-}\mathbf{N}_{jk}\mathbf{U}^{k-}
\label{clasaction}
\te
$\mathbf{N}$ is the noise self-correlation from \ref{eq1}.

\subsection{The CTP effective action}

Our goal is not the generating functional but rather the 2PI effective action $\Gamma$, which is the Legendre transform of the generating functional $W$ with respect to all sources. This means that we treat all sources, whether they are coupled to a single field or to a binary product, or else to a physical or to a mirror field, on the same footing. $\Gamma$ takes the form \cite{CalHu08}

\be
\Gamma_=S\left[\mathbf{\bar{u}}^J,\bar{p}^A\right]+S_2-\frac {ic}2\mathrm{Tr}\ln\left[\mathbf{G}\right]+\Gamma_Q
\te
where $\mathbf{G}$ denotes the matrix of all $64$ propagators

\be
\mathbf{G}=\left(
\begin{array}{cc}
G^{JK} & H^{JB}
\\
H^{AK} & I^{AB}
\end{array}
\right)
\label{matrixform}
\te
and
\bea
S_2&=&\frac12\left\{D_{jk}\left(G^{k+,j-}+G^{j-,k+}\right)\right.\nn
&+&2\mathbf{\bar{u}}^{j-}g_{jkl}G^{k+,l+}+2g_{jkl}\mathbf{\bar{u}}^{k+}\left(G^{l+,j-}+G^{j-,l+}\right)\nn
&+&\left.L_{ja}\left(H^{j-,a+} +H^{j+,a-}+H^{a+,j-} +H^{a-,j+}\right)+i\mathbf{N}_{jk}G^{k-,j-}\right\}
\tea
Formally $\Gamma_Q$ is the sum of all 2PI vacuum graphs with full propagators in the internal lines and a cubic vertex $g_{jkl}$ \cite{CalHu08}. Observe that since only $\mathbf{u}$ fields partake in the interaction, $\Gamma_Q$ is a functional of the $G^{JK}$ alone. Also that it is independent of the mean fields. This, and the fact that it is random Galilean invariant (RGI), as we shall see in next Section, are its two most important properties.

Let us summarize what we have said so far. Making a further and most extreme compresion of our notation, we may describe our model as a theory of fields $X^{\mu}$ with a classical action

\be
S=\frac12\left[\mathbf{D}_{\mu\nu}+i\mathbf{N}_{\mu\nu}\right]X^{\mu}X^{\nu}+\frac16g_{\mu\nu\rho}X^{\mu}X^{\nu}X^{\rho}
\te
We enlarge this theory by adding a source action

\be
S_J=J_{\mu}X^{\mu}+\frac12K_{\mu\nu}X^{\mu}X^{\nu}
\te
and define the generating functional

\be
e^{iW/c}=\int\:DX\:e^{i\left[S+S_J\right]/c}
\te
thereby introducing mean fields

\be
\frac{\delta W}{\delta J_{\mu}}=\bar{x}^{\mu}
\te
and propagators

\be
\frac{\delta W}{\delta K_{\mu\nu}}=\frac12\left[\bar{x}^{\mu}\bar{x}^{\nu}+G^{\mu\nu}\right]
\te
The full Legendre transform of the generating functional gives the 2PI effective action

\be
\Gamma =S\left[\bar{x}\right]+\frac12\left[\mathbf{D}_{\mu\nu}+i\mathbf{N}_{\mu\nu}+g_{\mu\nu\rho}\bar{x}^{\rho}\right]G^{\mu\nu}-\frac {ic}2\mathrm{Tr}\ln\left[{G}\right]+\Gamma_Q
\label{exact}
\te
where $\Gamma_Q$ is independent of the mean fields. 

It is a fundamental property of the Legendre transformation that it may be inverted. This leads to the equations of motion

\be
\frac{\delta \Gamma}{\delta \bar{x}^{\mu}}=\left[\mathbf{D}_{\mu\nu}+i\mathbf{N}_{\mu\nu}\right]\bar{x}^{\nu}+\frac12g_{\mu\nu\rho}\left[\bar{x}^{\nu}\bar{x}^{\rho}+G^{\nu\rho}\right]=-J_{\mu}-K_{\mu\nu}\bar{x}^{\nu}
\label{bfe}
\te

\be
\frac{\delta \Gamma}{\delta G^{\mu\nu}}=\frac12\left\{\mathbf{D}_{\mu\nu}+i\mathbf{N}_{\mu\nu}+g_{\mu\nu\rho}\bar{x}^{\rho}-icG^{-1}_{\mu\nu}- c\Sigma_{\mu\nu}\right\}=\frac{-1}2K_{\mu\nu}
\label{sde}
\te
where we have introduced the self energies

\be
\frac{\partial\Gamma_Q}{\partial G^{\mu\nu}}=\frac{-c}{2}\Sigma_{\mu\nu}
\te
We see that in this formulation the turbulence model becomes completely analogous to a quantum field theory problem \cite{CalHu08}, with $c$ playing the role of Planck's constant. In this sense, we may call $S$ the ``classical'' action, as oppossed to the ``quantum'' action $\Gamma$. Indeed if $c\mapsto 0$ the path integral is dominated by its saddle points, which means that the evolution of the mean fields is well described by the Navier-Stokes equations generated by $S$. The problem is that fully developed turbulence occurs in the opposite limit $c\mapsto\infty$.

\subsection{Physical solutions}
In the previous sections we have doubled the degrees of freedom of the theory by matching each physical field to a mirror field. We have at least two good reasons to do so, namely that in the enlarged theory the dynamical equations may be derived from the variation of an action functional, and that the relevant symmetry (which is random Galilean invariance) will be easier to work with in the enlarged theory. As we shall see in next section, the real part of the $S$ action (the part that is independent of the external forcing) is naturally RGI.

However, only some of the solutions of the enlarged theory really describe physical flows. To see this, let us use $\mathcal{J}$ for the string of all the sources. Then we have (assuming real sources)

\be
e^{-iW\left[\mathcal{J}\right]^*/c}=\int\:D\mathbf{U}DP\;\exp\left\{-i\left[S^*+S_J\right]/c\right\}
\label{genfun2}
\te
but, writing explicitly the branch indexes,

\be
-S\left[\mathbf{U}^{j\alpha},P^{a\beta}\right]^*=S\left[\alpha\mathbf{U}^{j\alpha},\beta P^{a\beta}\right]
\te
(no sum over $\alpha$ and $\beta$) so

\be
-W\left[J_{j\alpha},J_{a\beta},K_{j\alpha,k\beta},K_{j\alpha,a\beta},K_{a\alpha,b\beta}\right]^*=W\left[-\alpha J_{j\alpha},-\beta J_{a\beta},-\alpha\beta K_{j\alpha,k\beta},-\alpha\beta K_{j\alpha,a\beta},-\alpha\beta K_{a\alpha,b\beta}\right]
\te
Therefore at the physical point $\mathcal{J}=0$ we must have $\mathbf{\bar{u}}^{j-}=\bar{p}^{a-}=0$. $G^{j+,k+}$, $G^{j-,k-}$, $H^{j+,a+}$, $H^{j-,a-}$, $I^{a+,b+}$ and $I^{a-,b-}$ must be real, while $G^{j+,k-}$, $H^{j+,a-}$, $H^{j-,a+}$ and $I^{a+,b-}$ must be imaginary.

We can be more restrictive about the properties of meaningful solutions. Let us now look more closely into the generating functional $W\left[J^{U-},J^{P-}\right]$ which is obtained when all sources other than $J^{U-}$ and $J^{P-}$ are set to zero. From the definition \ref{fungen} we get

\be
e^{iW\left[J^{U-}\right]/c}=\int\:D\mathbf{f}D\mathbf{U}DP\;\mathcal{P}\left[\mathbf{f} \right]\delta\left(\mathbf{D}_j\left[\mathbf{U},P\right]-\mathbf{f}_j+J^{U-}_j\right)\delta\left(L^{\dagger}_{aj}\mathbf{U}^j+J^{P-}_a\right)
\te
Since the determinants of the operators involved are field independent, we get $W\left[J^{U-},J^{P-}\right]=\mathrm{constant}$. This shows that on a physical point we have the stronger constraints 

\be
\mathbf{\bar{u}}^{j-}=\bar{p}^{a-}=G^{j-,k-}=H^{j-,a-}=I^{a-,b-}=0
\te
Observe that while $\mathbf{\bar{u}}^{j-}$ and $G^{j-,k-}$ must vanish at a physical point, mixed propagators are meaningful. A case in point is the retarded propagator

\be
G^{j+,k-}=-ic\frac{\delta \mathbf{\bar{u}}^{j}}{\delta J_{k-}}\equiv -icG_{ret}
\te
which will play a leading role in what follows \cite{McCKiy05}.

\section{Random Galilean invariance}
As we have already mentioned, important features of our model may be linked to the way it is affected by random Galilean transformations \cite{Kra65,HorLip79,Spe88,MCCO94,BLPP98,BerHoc07}. In this section we introduce these transformations and investigate the transformation properties of both the ``classical'' action $S$ and the ``quantum'' action $\Gamma$.

\subsection{Classical random Galilean invariance}
We shall now introduce a class of transformations that leaves the real part of the ``classical'' action \ref{clasaction} invariant. Let $\mathbf{E}^p\left(t\right)$ be a time dependent homogeneous vector field, and

\be
\mathbf{y}^p\left(t\right)=\int_{-\infty}^t\:d\tau\:\mathbf{E}^p\left(\tau\right)
\te
Now define the transformation
\bea
\mathbf{U}^{p+}\left(\mathbf{x},t\right)&\mapsto&\mathbf{E}^p\left(t\right)+\mathbf{U}^{p+}\left(\mathbf{x}-\mathbf{y}\left(t\right),t\right)\nn
&=&\mathbf{U}^{p+}\left(\mathbf{x},t\right)+\mathbf{E}^p\left(t\right)-\int_{-\infty}^{\infty}\:d\tau\:\mathbf{E}^q\left(\tau\right)\theta \left(t-\tau\right)\nabla_q\mathbf{U}^{p+}\left(\mathbf{x},t\right)+O\left(E^2\right)
\tea

\bea
\mathbf{U}^{p-}\left(\mathbf{x},t\right)&\mapsto&\mathbf{U}^{p-}\left(\mathbf{x}-\mathbf{y}\left(t\right),t\right)\nn
&=&\mathbf{U}^{p-}\left(\mathbf{x},t\right)-\int_{-\infty}^{\infty}\:d\tau\:\mathbf{E}^q\left(\tau\right)\theta \left(t-\tau\right)\nabla_q\mathbf{U}^{r-}\left(\mathbf{x},t\right)+O\left(E^2\right)
\tea

\bea
P^+\left(\mathbf{x},t\right)&&\mapsto -\dot{\mathbf{E}}^p\left(t\right)\left[\mathbf{x}_p-\mathbf{y}_p\left(t\right)
\right]+P^{+}\left(\mathbf{x}-\mathbf{y}\left(t\right),t\right)\nn
&=&P^+\left(\mathbf{x},t\right)-\dot{\mathbf{E}}^p\left(t\right)\mathbf{x}_p-\int_{-\infty}^{\infty}\:d\tau\:\mathbf{E}^q\left(\tau\right)\theta \left(t-\tau\right)\nabla_qP^+\left(\mathbf{x},t\right)+O\left(E^2\right)
\tea

\bea
P^-\left(\mathbf{x},t\right)&\mapsto&\mathbf{E}_p\left(t\right)\mathbf{U}^{p-}\left(\mathbf{x}-\mathbf{y}\left(t\right),t\right)
+P^{-}\left(\mathbf{x}-\mathbf{y}\left(t\right),t\right)\nn
&=&P^-\left(\mathbf{x},t\right)+\mathbf{E}_p\left(t\right)\mathbf{U}^{p-}\left(\mathbf{x},t\right)-\int_{-\infty}^{\infty}\:d\tau\:\mathbf{E}^q\left(\tau\right)\theta \left(t-\tau\right)\nabla_qP^-\left(\mathbf{x},t\right)+O\left(E^2\right)
\tea

We obtain the transformation rules

\be
\nabla_p{\mathbf{U}}^{p\pm}\left(\mathbf{x},t\right)\mapsto\nabla_p{\mathbf{U}}^{p\pm}\left(\mathbf{x}-\mathbf{y}\left(t\right),t\right)
\te

\be
\nabla_p{P}^{+}\left(\mathbf{x},t\right)\mapsto -\dot{\mathbf{E}}_p\left(t\right)+\nabla_p{P}^{+}\left(\mathbf{x}-\mathbf{y}\left(t\right),t\right)
\te

\be
\nabla_p{P}^{-}\left(\mathbf{x},t\right)\mapsto\nabla_p\left[\mathbf{E}_q\mathbf{U}^{q-}\left(\mathbf{x}-\mathbf{y}\left(t\right),t\right)\right]+\nabla_p{P}^{-}\left(\mathbf{x}-\mathbf{y}\left(t\right),t\right)
\te
and

\be
\frac{\partial}{\partial t}{\mathbf{U}}^{p+}\left(\mathbf{x},t\right)\mapsto\dot{\mathbf{E}}^p\left(t\right)+\frac{\partial}{\partial t}\mathbf{U}^{p+}\left(\mathbf{x}-\mathbf{y}\left(t\right),t\right)-\mathbf{E}^q\left(t\right)\nabla_q{\mathbf{U}}^{p+}\left(\mathbf{x}-\mathbf{y}\left(t\right),t\right)
\te
This transformation turns the action into 

\be
S\left[{\mathbf{U}},{P}\right]\mapsto S\left[\mathbf{U},P\right]+
\frac i2\int\:d\mathbf{x}dt\:d\mathbf{x'}dt'\:\delta_E\mathbf{N}_{pq}\left[\mathbf{x},t;\mathbf{x'},t'\right]{\mathbf{U}}^{p-}\left(\mathbf{x},t\right){\mathbf{U}}^{q-}\left(\mathbf{x'},t'\right)
\te
where

\be
\delta_E\mathbf{N}_{pq}\left[\mathbf{x},t;\mathbf{x'},t'\right]=\mathbf{N}_{pq}\left[\mathbf{x}+\mathbf{y}\left(t\right),t;\mathbf{x'}+\mathbf{y}\left(t'\right),t'\right]-\mathbf{N}_{pq}\left[\mathbf{x},t;\mathbf{x'},t'\right]
\te
Observe that $\delta_E\mathbf{N}_{pq}=0$ when the noise self-correlation is of the form \ref{eq2}.

\subsection{``Quantum'' random galilean invariance}
Now we discuss random galilean invariance of the full ``quantum'' action $\Gamma$. 

Let us summarize the results from the previous subsection in terms of the most compressed notation introduced in Section II. As we have shown, there is a class of transformations

\be
X^{\mu}\mapsto X^{\mu}+C^{\mu}_{\alpha}E^{\alpha}+F^{\mu}_{\nu\alpha}X^{\nu}E^{\alpha}+O\left(E^2\right)
\te
where the $E^{\alpha}$ are arbitrary parameters, such that the real part of the classical action is invariant and the imaginary part preserves its form

\be
S_C\mapsto S_C+\frac i2\delta\mathbf{N}_{\mu\nu}X^{\mu}X^{\nu}+O\left(E^2\right)
\te 
Observe that this implies the identities

\be
\left\{\mathbf{D}_{\mu\nu}X^{\nu}+\frac12g_{\mu\nu\rho}X^{\nu}X^{\rho}\right\}\left[C^{\mu}_{\alpha}+F^{\mu}_{\sigma\alpha}
X^{\sigma}\right]=0
\te
which may be decomposed into

\be
\mathbf{D}_{\mu\nu}C^{\mu}_{\alpha}=0
\te

\be
\mathbf{D}_{\mu\nu}F^{\mu}_{\rho\alpha}+\mathbf{D}_{\rho\nu}F^{\mu}_{\nu\alpha}+g_{\mu\nu\rho}C^{\mu}_{\alpha}=0
\label{ZJ9}
\te

\be
g_{\mu\left(\nu\rho\right.}F^{\mu}_{\left.\sigma\right)\alpha}=0
\label{ZJ10}
\te
and

\be
\mathbf{N}_{\mu\nu}C^{\mu}_{\alpha}=0
\te
We define
\be
\delta\mathbf{N}_{\mu\nu}=\delta\mathbf{N}_{\mu\nu\alpha}E^{\alpha}=2\mathbf{N}_{\mu\rho}F^{\rho}_{\nu\alpha}E^{\alpha}
\te
We also assume

\be
F^{\nu}_{\nu\alpha}=0
\te
making the measure of integration also invariant. It is easy to show that the actual random Galilean transformations defined above satisfy this condition.

The source action transforms into

\be
S_J\mapsto S_J+\left\{J_{\mu}C^{\mu}_{\alpha}+\left[J_{\nu}F^{\nu}_{\mu\alpha}+K_{\mu\nu}C^{\nu}_{\alpha}\right]X^{\mu}+\frac12\left[K_{\mu\rho}F^{\rho}_{\nu\alpha}+K_{\rho\nu}F^{\rho}_{\mu\alpha}\right]X^{\mu}X^{\nu}\right\}E^{\alpha}+O\left(E^2\right)
\te
The generating functional itself is invariant, and this leads to the identity

\be
\left\langle \frac{\delta S}{\delta E^{\alpha}}\right\rangle=0
\te
where

\be
\left\langle \frac{\delta S}{\delta E^{\alpha}}\right\rangle=J_{\mu}C^{\mu}_{\alpha}+\left[J_{\nu}F^{\nu}_{\mu\alpha}+K_{\mu\nu}C^{\nu}_{\alpha}\right]\bar{x}^{\mu}+\frac12\left[K_{\mu\rho}F^{\rho}_{\nu\alpha}+K_{\rho\nu}F^{\rho}_{\mu\alpha}+i\delta\mathbf{N}_{\mu\nu\alpha}\right]\left[\bar{x}^{\mu}\bar{x}^{\nu}+G^{\mu\nu}\right]
\te
We further eliminate the sources in terms of derivatives of the effective action

\be
0=-\frac{\delta \Gamma}{\delta \bar{x}^{\mu}}\left[C^{\mu}_{\alpha}+F^{\mu}_{\nu\alpha}\bar{x}^{\nu}\right]+\frac i2\delta\mathbf{N}_{\mu\nu}\bar{x}^{\mu}\bar{x}^{\nu}+\left[K_{\mu\rho}F^{\rho}_{\nu\alpha}+\frac i2\delta\mathbf{N}_{\mu\nu\alpha}
\right]G^{\mu\nu}
\te
Finally we obtain the Zinn-Justin identity \cite{CalHu08,Zin93}

\be
0=-\frac{\delta \Gamma}{\delta \bar{x}^{\mu}}\left[C^{\mu}_{\alpha}+F^{\mu}_{\nu\alpha}\bar{x}^{\nu}\right]-2\frac{\delta \Gamma}{\delta G^{\mu\rho}}F^{\rho}_{\nu\alpha}G^{\mu\nu}+ i\mathbf{N}_{\mu\rho}F^{\rho}_{\nu\alpha}\left[\bar{x}^{\mu}\bar{x}^{\nu}+G^{\mu\nu}\right]
\label{ZJ}
\te
If we use the explicit form \ref{exact} and the known transformation properties of the classical action, we get

\be
0=-\frac12g_{\mu\sigma\rho}G^{\sigma\rho}\left[C^{\mu}_{\alpha}+F^{\mu}_{\nu\alpha}\bar{x}^{\nu}\right]-\left[\mathbf{D}_{\mu\rho}+i\mathbf{N}_{\mu\rho}+g_{\mu\sigma\rho}\bar{x}^{\sigma}- icG^{-1}_{\mu\rho}- c\Sigma_{\mu\rho}\right]F^{\rho}_{\nu\alpha}G^{\mu\nu}+ i\mathbf{N}_{\mu\rho}F^{\rho}_{\nu\alpha}G^{\mu\nu}
\label{ZJa}
\te
Using the tracelessness of $F^{\rho}_{\nu\alpha}$ this reduces to

\be
0=-\frac12g_{\mu\sigma\rho}G^{\sigma\rho}\left[C^{\mu}_{\alpha}+F^{\mu}_{\nu\alpha}\bar{x}^{\nu}\right]-\left[\mathbf{D}_{\mu\rho}+g_{\mu\sigma\rho}\bar{x}^{\sigma}-c\Sigma_{\mu\rho}\right]F^{\rho}_{\nu\alpha}G^{\mu\nu}
\label{ZJb}
\te
Now use \ref{ZJ9}  

\be
0=\frac12g_{\mu\sigma\rho}G^{\sigma\rho}F^{\mu}_{\nu\alpha}\bar{x}^{\nu}+\left[g_{\mu\sigma\rho}\bar{x}^{\sigma}- c\Sigma_{\mu\rho}\right]F^{\rho}_{\nu\alpha}G^{\mu\nu}
\label{ZJe}
\te
and \ref{ZJ10}

\be
0=c\Sigma_{\mu\rho}F^{\rho}_{\nu\alpha}G^{\mu\nu}
\label{ZJc}
\te
This equation shows that the quantum part of the action is RGI. When the noise self correlation is of the form \ref{eq2} the full quantum action $\Gamma$ is RGI.

\subsection{Random galilean invariance and self-energies}
We have shown that the quantum part of the effective action is random galilean invariant. This suggests it can be written as a functional of the random galilean invariant part of the two point functions only.

Observe that a random galilean invariant kernel $F\left[\mathbf{x},t_x;\mathbf{y},t_y\right]$ must be of the form $f\left(\mathbf{x-y}\right)\delta\left(t_x-t_y\right)$. Therefore we may regard

\be
F_{\mathbf{z},t}\left[\mathbf{x},t_x;\mathbf{y},t_y\right]=\delta\left(\mathbf{x-y-z}\right)\delta\left(t_x-t\right)
\delta\left(t_y-t\right)
\te
as a basis of random galilean invariant kernels. We now seek an inner product where this basis is orthonormal, namely

\be
\left\langle F_{\mathbf{z},t},F_{\mathbf{z'},t'}\right\rangle=\delta\left(\mathbf{z-z'}\right)\delta\left(t-t'\right)
\te
The simplest solution is

\be
\left\langle F,G\right\rangle=\frac1{\Omega L^d}\int\:d\mathbf{x}dt_xd\mathbf{y}dt_y\:F\left[\mathbf{x},t_x;\mathbf{y},t_y\right]^*G\left[\mathbf{x},t_x;\mathbf{y},t_y\right]
\te
where $L^d$ is the available volume and we have defined $\delta\left(t-t'\right)^2=\Omega \delta\left(t-t'\right)$. Now we can define a projector on the space of random galilean invariant kernels

\be
G\mapsto G_{RGI}
\te
where

\be
G_{RGI}\left[G\right]=\int d\mathbf{z}dt\:F_{\mathbf{z},t}\left\langle F_{\mathbf{z},t},G\right\rangle
\te

If $\Gamma_Q$ is really a functional of $G_{RGI}$ rather than simply $G$, then it follows that

\be
\Sigma_{r\alpha,s\beta}\left[\mathbf{x},t_x;\mathbf{y},t_y\right]=\tilde\Sigma_{r\alpha,s\beta}\left[\mathbf{x-y},t_x\right]\delta\left(t_x-t_y\right)
\label{rgise}
\te
In the particular case of a homogeneous solution, we may drop the $t_x$ dependence of $\tilde\Sigma_{r\alpha,s\beta}$. 

We see that random Galilean invariance puts  very heavy constraints on the structure of the self-energies. In Section IV we shall use these constraints to investigate the velocity correlations in fully developed turbulence.

\section{The structure of the 2PI equations}
From this point on, we shall investigate solutions where all background fields vanish. Therefore, we may disregard the background field equations \ref{bfe}, except for the constraints

\be
\left.\nabla_qG^{p+,q+}\left[\mathbf{x},t;\mathbf{y},t_y\right]\right|_{\mathbf{x}=\mathbf{y},t=t_y}=\left.\nabla_qG^{p-,q+}\left[\mathbf{x},t;\mathbf{y},t_y\right]\right|_{\mathbf{x}=\mathbf{y},t=t_y}=0
\te
The dynamics of the model is determined by the Schwinger-Dyson equations \ref{sde}, where moreover we put all sources equal to zero.

Our strategy relies on the observation that we have two different ways of computing the inverse correlations $\left[\mathbf{G}^{-1}\right]_{\mu\nu}$. On one hand, the equations of motion imply

\be
\left[\mathbf{G}^{-1}\right]_{\mu\nu}=\left(\frac{-i}c\right)\left[\mathbf{D}_{\mu\nu}+i\mathbf{N}_{\mu\nu}-c\Sigma_{\mu\nu}\right]
\te
If we split the velocity and pressure fields, this means

\be
\left[\mathbf{G}^{-1}\right]_{\mu\nu}=\left(\frac{-i}c\right)\left(
\begin{array}{cc}
\mathcal{D}_{JK} & L_{JB}
\\
L_{AK} & 0
\end{array}
\right)
\label{invmatrixform}
\te
where

\be
\mathcal{D}_{JK}=D_{JK}+iN_{JK}-c\Sigma_{JK}=\left(\begin{array}{cc}
-c\Sigma_{j+k+} & D_{j+k-}-c\Sigma_{j+k-}
\\
D_{j-k+}-c\Sigma_{j-k-} & iN_{j-k-}-c\Sigma_{j-k-}
\end{array}
\right)
\label{invmatrixform3}
\te
On the other hand, we have both 

\be
\left[\mathbf{G}^{-1}\right]_{\mu\nu}\mathbf{G}^{\nu\rho}=\delta_{\mu}^{\rho}
\te
and

\be
\mathbf{G}^{\rho\nu}\left[\mathbf{G}^{-1}\right]_{\nu\mu}=\delta_{\mu}^{\rho}
\te
From the representation \ref{matrixform}, we immediately find

\be
\frac{-i}c L_{AK}H^{KB}=\delta^B_A
\te
Writing this in full with the help of \ref{eq8} we get

\be
\nabla_pH^{p\alpha,\beta}\left[\mathbf{x},t;\mathbf{y},t_y\right]=-ic\sigma^{\alpha\beta}\delta\left(\mathbf{x}
-\mathbf{y}\right)\delta\left(t-t_y\right)
\label{eq22}
\te
where $\sigma$ is the first Pauli matrix

\be
\sigma=\left(
\begin{array}{cc}
0 & 1
\\
1 & 0
\end{array}
\right)
\label{matrixform2}
\te
which here plays the role of a ``metric tensor'' for the branch indexes. The point of \ref{eq22}  is that it has no dynamical information, though the solution may depend on the boundary conditions on the flow. For flows in unbound space, as we are considering here, we may write down the solution right away

\be
H^{p\alpha,\beta}\left[\mathbf{x},t;\mathbf{y},t_y\right]=\int\:\frac{d\mathbf{k}}{\left(2\pi\right)^d}\:\frac{d\omega}{\left(2\pi\right)}\:e^{i\left[\mathbf{k}\left(\mathbf{x}-\mathbf{x'}\right)-\omega\left(t-t'\right)\right]}H^{p\alpha,\beta}\left[\mathbf{k},\omega\right]
\label{ft}
\te

\be
H^{p\alpha,\beta}\left[\mathbf{k},\omega\right]=-c\frac{\mathbf{k}^p}{k^2}\sigma^{\alpha\beta}
\te
similarly we get

\be
\nabla_pG^{p\alpha,q\beta}\left[\mathbf{x},t;\mathbf{y},t_y\right]=0
\label{eq23}
\te
which means that all possible $36$ velocity correlators are transverse.

The velocity correlations are determined by the equations

\be
\mathcal{D}_{JK}G^{KL}=ic\Delta^L_J
\te
where

\be
\Delta^L_J=\delta^L_J+\frac icL_{JB}H^{BL}\equiv \delta^{\beta}_{\alpha}\Delta^l_j
\te
This operator admits a Fourier transform analogous to \ref{ft}, with kernel

\be
\Delta^{q}_p=\delta^{q}_p-\frac{\mathbf{k}_p\mathbf{k}^q}{k^2}
\te
For a physical solution, we must have $G^{p-,q-}=0$. Therefore

\be
-c\Sigma_{j+,k+}G^{k+,l-}=0
\te

\be
\left[D_{j-,k+}-c\Sigma_{j-,k+}\right]G^{k+,l-}=ic\Delta^l_j
\te
Since the second equation says that $G^{k+,l-}$ is regular, the first implies that  $\Sigma_{j+,k+}=0$. 

The final two equations now read

\be
\left[D_{j+,k-}-c\Sigma_{j+,k-}\right]G^{k-,l+}=ic\Delta^l_j
\te

\be
\left[D_{j-,k+}-c\Sigma_{j-,k+}\right]G^{k+,l+}+\left[i\mathbf{N}_{jk}-c\Sigma_{j-,k-}\right]G^{k-,l+}=0
\te
To extract the meaning of these equations, let us write

\be
\mathcal{D}_{p\alpha,q\beta}\left[\mathbf{x},t;\mathbf{y},t_y\right]=\bar{\mathcal{D}}_{p\alpha,q\beta}+\nabla^{x}_pA_{q,\alpha,\beta}+
\nabla^{y}_qB_{p,\alpha,\beta}+\nabla^{x}_p\nabla^{y}_qC_{\alpha,\beta}
\te
where

\be
\nabla^{y}_q\bar{\mathcal{D}}_{p\alpha,q\beta}=\nabla^{x}_p\bar{\mathcal{D}}_{p\alpha,q\beta}=\nabla^{y}_qA_{q,\alpha,\beta}=\nabla^{x}_pB_{p,\alpha,\beta}=0
\te
Then the above equations decompose into 

\be
A_{q,\alpha,\beta}=B_{p,\alpha,\beta}=0
\te

\be
\bar{\mathcal{D}}_{j-,k+}G^{k+,l-}=ic\Delta^l_j
\label{causaleq}
\te

\be
\bar{\mathcal{D}}_{j+,k-}G^{k-,l+}=ic\Delta^l_j
\te

\be
\bar{\mathcal{D}}_{j-,k+}G^{k+,l+}+\bar{\mathcal{D}}_{j-,k-}G^{k-,l+}=0
\te
which admits the formal solution

\be
G^{k+,l+}=\frac icG^{k+,j-}\bar{\mathcal{D}}_{j-,k-}G^{k-,l+}
\te
or else, defining

\be
G^{k+,j-}=-icG^{kj}_{ret}
\te

\be
G^{k-,l+}=-icG^{kl}_{adv}
\te

\be
\bar{\mathcal{D}}_{j-,k-}=i\mathcal{N}_{jk}
\label{mathcaln}
\te
the physical velocity correlation

\be
G^{k+,l+}=cG^{kj}_{ret}\mathcal{N}_{jk}G^{kl}_{adv}
\label{fdt}
\te
which may be regarded as the fluctuation-dissipation theorem for homogeneous turbulence \cite{McCKiy05}. The velocity fluctuations described by $G^{++}$ are equivalent to the stochastic fluctuations of a fluid under a renormalized force $\mathbf{f}^{ren}_j$ with self correlation

\be
\left\langle \mathbf{f}^{ren}_j\mathbf{f}^{ren}_k\right\rangle=c\mathcal{N}_{jk}
\te

Analog considerations may be done for the pressure-pressure correlations. In this case the relevant equation is

\be
\mathcal{D}_{JK}H^{KB}+L_{JA}I^{AB}=0
\te
linking the pressure-pressure correlations to the longitudinal part of the self-energies.

\subsection{Fully developed turbulence}
We have already introduced the macroscopic length scale $L$ and velocity scale $V$, in terms of which we have the constant $c=(L^dV^2)$. We say that turbulence is fully developed when the statistical properties of the flow are determined by the scales $\epsilon =V^3L^{-1}$ and $\nu_b =LV/\mathrm{Re}$ alone, where $\mathrm{Re}$ is Reynolds' number. Alternatively, we may introduce the microscopic length and velocity scales

\be
\lambda =\left(\frac{\nu_b^3}{\epsilon}\right)^{1/4}=\frac L{\mathrm{Re}^{3/4}}
\te

\be
u^*=\left(\nu_b\epsilon\right)^{1/4}=\frac V{\mathrm{Re}^{1/4}}
\te
In terms of these scales we have

\be
c={L^dV^2}={\mathrm{Re}^{3d+2/4}}{\lambda^du^{*2}}
\te
The limit $\mathrm{Re}\mapsto 0$ with $\lambda$ and $u^*$ fixed corresponds to laminar flow. In this limit $c\mapsto 0$ and the path integral is dominated by its stationary points. In the opposite limit $c\mapsto \infty$ and there is no simple way of computing the path integral.

In the inertial range, the flow becomes independent of the viscosity $\nu_b$. We shall take as an experimental fact that many properties, such as equal time two and three point functions, are ``universal'', meaning that they  are determined by the single dimensionfull quantity $\epsilon$ \cite{MonYag71}. Moreover, we shall also accept as an experimental fact that for two and three point correlations those properties can be derived by simple dimensional analysis. Here ``simple'' means that we disregard anomalous dimensions, which may be important for higher correlations \cite{BLPP98,FRIS95,MarMin09}.

Let us assume a traslation invariant, steady flow where all background fields vanish. Then the $++$ propagators can be written in terms of their Fourier transforms

\be
G^{++}\left(\mathbf{x},t;\mathbf{x'},t'\right)=\int\:\frac{d\mathbf{k}}{\left(2\pi\right)^d}\:\frac{d\omega}{\left(2\pi\right)}\:e^{i\left[\mathbf{k}\left(\mathbf{x}-\mathbf{x'}\right)-\omega\left(t-t'\right)\right]}G^{++}\left[\mathbf{k},\omega\right]
\label{Fourier}
\te
The equal time correlations have Fourier transforms

\be
G^{++}_{eq}\left[\mathbf{k}\right]=\int\:\frac{d\omega}{\left(2\pi\right)}\:G^{++}\left[\mathbf{k},\omega\right]
\te
The dimensions of $G^{++}_{eq}$ are $k^{-d}V^2$, and since the only velocity scale that can be built out of $\epsilon$ and $k$ is $\left(\epsilon /k\right)^{1/3}$ we find the Kolmogorov spectrum

\be
G^{++}_{eq}\left[\mathbf{k}\right]\sim C_K\epsilon^{2/3}k^{-d-\left(2/3\right)}
\te

We shall now extend this analysis to all the velocity correlations and the corresponding self-energies. We assume the flow is homogeneous and isotropic. This means all correlations can be written in terms of their Fourier transforms as in \ref{Fourier}. Moreover we can transversality explicit by writing

\be
G^{p\alpha,q\beta}=c\Delta^{pq}\mathcal{G}^{\alpha,\beta}
\te

Let us recall that $G^{p\alpha,q\beta}\left[\mathbf{k},\omega\right]$ has units of $V^2k^{-d}\omega^{-1}$ and therefore
$\mathcal{G}^{\alpha,\beta}\left[k,\omega\right]\sim cG^{r\alpha,s\beta}\left[\mathbf{k},\omega\right]$ has units of $\omega^{-1}$. 

Since the self-energy $\Sigma_{j-,k+}$ has the structure  \ref{rgise}, its Fourier transform

\be
\Sigma_{p-,q+}\left[\mathbf{k},\omega\right]=\frac1c\left\{\Delta_{pq}\left[\nu\left[k\right]+i\Omega\left[k\right]\right]+\frac{\mathbf{k}_p\mathbf{k}_q}{k^2}\mathcal{L}\left[k\right]\right\}
\te
It follows that the equation for $\mathcal{G}_{ret}=i\mathcal{G}^{+-}$, in the limit $\nu_b\mapsto 0$, is (cfr. \ref{causaleq})

\be
\left[i\left(\omega-\Omega\left[k\right]\right)-\nu\left[k\right]\right]\mathcal{G}_{ret}=-1
\label{causaleq2}
\te
Recall that $\mathcal{G}_{adv}=\mathcal{G}_{ret}^*$.

Since the retarded propagator is real and causal, we must have $\Omega =0$ and $\nu\ge 0$. Moreover $\nu$ has dimensions of frequency, and therefore we must have

\be
\nu\left[k\right]=\nu_0\left(k^2\epsilon\right)^{1/3}
\te
where $\nu_0$ is a dimensionless constant. 

We have already shown that the velocity fluctuations described by $G^{++}$ are equivalent to the stochastic fluctuations of a fluid under a renormalized force $\mathbf{f}^{ren}_j$ with self correlation $\mathcal{N}$ given in \ref{mathcaln}. If the bare noise self correlation $\mathbf{N}$ has the random galilean invariant form \ref{eq2}, then the same will be true of the renormalized noise correlation $\mathcal{N}$. Projecting over the transverse part, we can write \ref{fdt} as

\be
\mathcal{G}^{++}\left[k,\omega\right]=c\;\frac{\mathcal{N}\left[k\right]}{\omega^2 +\nu_0^2\left(k^2\epsilon\right)^{2/3}}
\te
Direct integration yields

\be
G^{++}_{eq}\left[\mathbf{k}\right]=\frac c{2\nu_0}\frac{\mathcal{N}\left[k\right]}{\left(k^2\epsilon\right)^{1/3}}
\te
and so the Kolmogorov spectrum implies the well known result

\be
\mathcal{N}\left[k\right]= N_0 k^{-d}
\te
where

\be
N_0=\frac{2\nu_0}cC_K\epsilon
\te

\section{Kadanoff-Baym equations for nearly homogeneous and isotropic flows}

We now have all the necessary elements to consider the regression to homogeneity and isotropy in high Reynolds number flow. We have in mind a situation where there are no mean flows and the fluctuations are already nearly traslation invariant. In such a situation it is convenient to introduced centroid coordinate and time \cite{KadBay62,{EdwMcC72},CalHu88,CalHu08} $\mathbf{X}=\left(1/2\right)\left(\mathbf{x+y}\right)$ and $T=\left(1/2\right)\left(t_x+t_y\right)$ and write

\be
G^{p\beta,q\gamma}\left[\mathbf{x},t_x;\mathbf{y},t_y\right]=\int\frac{d\mathbf{k}}{\left(2\pi\right)^d}
\frac{d\omega}{\left(2\pi\right)}\:e^{i\left[\mathbf{k\left(x-y\right)}-\omega\left(t_x-t_y\right)\right]}G^{p\beta,q\gamma}\left[\mathbf{k},\omega,\mathbf{X},T\right]
\te
to obtain the partial Fourier transform of $\left[G^{-1}\right]_{\mu\nu}$ from that of $G^{\nu\rho}$. Of course, we cannot do this exactly; we shall be content to find the \emph{adiabatic expansion} of $\left[G^{-1}\right]_{\mu\nu}$, meaning an expansion where the different terms are classified according to the number of derivatives with respect to the centroid variables. We can also derive the inverse propagators from the variation of the 2PI CTP EA. Demanding that both computations of the inverse propagators yield the same result becomes a set of equations for the partial Fourier transforms of the propagators. Truncating these equations to first adiabatic order yields the so-called Kadanoff-Baym equations.

The pressure-pressure corelations are slaved to the longitudinal part of the inverse velocity-velocity correlations and we shall not consider them further. We also have the constraint

\be
\int\frac{d\mathbf{k}}{\left(2\pi\right)^d}
\frac{d\omega}{\left(2\pi\right)}\left[i\mathbf{k}_p+\frac12\frac{\partial}{\partial\mathbf{X}^p}\right] G^{p+,q+}\left[\mathbf{k},\omega,\mathbf{X},T\right]=0
\te
which means that the velocity correlations do not induce a nonzero mean velocity. 

As a matter of fact, the analysis of the first half of Section IV holds, because we did not assume homogeneity there. The only new element is that we must derive an expression for the adiabatic expansion of the composition of two nearly homogeneous kernels, and we have to see what constraints are necessary to enforce the transversality of the velocity correlations, \ref{eq23}.

Consider two kernels

\be
F\left[\mathbf{x},t_x;\mathbf{y},t_y\right]=\int\frac{d\mathbf{k}}{\left(2\pi\right)^d}
\frac{d\omega}{\left(2\pi\right)}\:e^{i\left[\mathbf{k\left(x-y\right)}-\omega\left(t_x-t_y\right)\right]}F\left[\mathbf{k},\omega,\mathbf{X},T\right]
\te
and

\be
G\left[\mathbf{x},t_x;\mathbf{y},t_y\right]=\int\frac{d\mathbf{k}}{\left(2\pi\right)^d}
\frac{d\omega}{\left(2\pi\right)}\:e^{i\left[\mathbf{k\left(x-y\right)}-\omega\left(t_x-t_y\right)\right]}G\left[\mathbf{k},\omega,\mathbf{X},T\right]
\te
Write

\bea
F\ast G\left[\mathbf{x},t_x;\mathbf{y},t_y\right]&=&\int d\mathbf{z}dt_z\:F\left[\mathbf{x},t_x;\mathbf{z},t_z\right]G\left[\mathbf{z},t_z;\mathbf{y},t_y\right]\nn
&=&\int\frac{d\mathbf{k}}{\left(2\pi\right)^d}
\frac{d\omega}{\left(2\pi\right)}\:e^{i\left[\mathbf{k\left(x-y\right)}-\omega\left(t_x-t_y\right)\right]}\left(F*G\right)\left[\mathbf{k},\omega,\mathbf{X},T\right]
\tea
The required formula is \cite{CalHu08}

\be
\left(F*G\right)\left[\mathbf{k},\omega,\mathbf{X},T\right]=FG-\frac i2\left\{F,G\right\}
\te
where

\be
\left\{F,G\right\}=\nabla_kF\nabla_XG-\frac{\partial F}{\partial\omega}\frac{\partial G}{\partial T}-\left(F\leftrightarrow G\right)
\te
Let us now consider the transversality constraint.
In the strictly translation invariant case, the Fourier transform $G^{r\alpha,s\beta}\left[\mathbf{k},\omega\right]$ of the propagators can be written as a linear combination of the two tensors $\Delta^{rs}$ and $\mathbf{k}^r\mathbf{k}^s/k^2$. The transversality condition enforces the vanishing of the latter term. To first adiabatic order, we may add a new independent vector as the gradiant with respect of the centroid variable of a scalar function. Projecting out the transverse part of such a vector, we may write

\be
G^{r\alpha,s\beta}=c\left\{\Delta^{rs}\mathcal{G}^{\alpha,\beta}+\frac{\mathbf{k}^r\mathbf{k}^s}{k^2}\mathcal{G}^{\alpha,\beta}_1+\frac{\mathbf{k}^r}{k^2}\Delta^{ts}\mathcal{G}^{\alpha,\beta}_{2t}+\Delta^{rt}\frac{\mathbf{k}^s}{k^2}\mathcal{G}^{\alpha,\beta}_{3t}\right\}
\te
where $\mathcal{G}^{\alpha,\beta}_1$, $\mathcal{G}^{\alpha,\beta}_{2t}$ and $\mathcal{G}^{\alpha,\beta}_{3t}$ are quantities of first adiabatic order. The transversality condition

\be
\left[i\mathbf{k}_r+\frac12\frac{\partial}{\partial\mathbf{X}^r}\right] G^{r\alpha,s\beta}\left[\mathbf{k},\omega,\mathbf{X},T\right]=0
\te
becomes

\be
0=\Delta^{rs}\frac12\frac{\partial}{\partial\mathbf{X}^r}\mathcal{G}^{\alpha,\beta}+i{\mathbf{k}^s}\mathcal{G}^{\alpha,\beta}_1+i\Delta^{rs}\mathcal{G}^{\alpha,\beta}_{2r}
\te
so

\be
\mathcal{G}^{\alpha,\beta}_1=0
\te
and

\be
\mathcal{G}^{\alpha,\beta}_{2r}=\frac i2\frac{\partial}{\partial\mathbf{X}^r}\mathcal{G}^{\alpha,\beta}
\te
The second transversality condition

\be
\left[-i\mathbf{k}_s+\frac12\frac{\partial}{\partial\mathbf{X}^s}\right] G^{r\alpha,s\beta}\left[\mathbf{k},\omega,\mathbf{X},T\right]=0
\te
implies

\be
\mathcal{G}^{\alpha,\beta}_{3r}=-\mathcal{G}^{\alpha,\beta}_{2r}
\te
Let us make a corresponding expansion for the inverse propagators

\be
\mathcal{D}_{p\alpha,q\beta}=\Delta_{rs}\bar{\mathcal{D}}_{\alpha,\beta}+\frac{\mathbf{k}_r\mathbf{k}_s}{k^2}\mathcal{D}^{long}_{\alpha,\beta}+\frac{i}{2k^2}\left[\mathbf{k}^r\Delta^{ts}
-\Delta^{rt}\mathbf{k}^s\right]
\mathcal{D}_{2t\alpha,\beta}
\te
Then we get

\be
\bar{\mathcal{D}}_{-,+}\mathcal{G}^{+,-}-\frac i2\left\{\bar{\mathcal{D}}_{-,+},\mathcal{G}^{+,-}\right\}=ic
\te

\be
\mathcal{D}_{2u-,+}\mathcal{G}^{+,-}+\bar{\mathcal{D}}_{-,+}\frac{\partial}{\partial\mathbf{X}^u}\mathcal{G}^{+,-}-\frac{\partial}{\partial\mathbf{X}^u}\mathcal{D}^{long}_{-,+}\mathcal{G}^{+,-}=0
\te

\be
\frac{\partial}{\partial\mathbf{X}^t}\left[\bar{\mathcal{D}}_{-,+}\mathcal{G}^{+,-}\right]=0
\te
If we introduce the retarded propagator

\be
\mathcal{G}^{+,-}={-i}\mathcal{G}_{ret}
\te
then

\be
\bar{\mathcal{D}}_{-,+}=-\left[\mathcal{G}_{ret}\right]^{-1}
\te

\be
\mathcal{D}_{2u-,+}=\frac{\partial}{\partial\mathbf{X}^u}\left[\bar{\mathcal{D}}_{-,+}+\mathcal{D}^{long}_{-,+}\right]
\te
We are almost done. Folowing \cite{CalHu08} we define the density of states

\be
\mathcal{D}=\frac1\pi\:\mathrm{Im}\left[\mathcal{G}_{ret}\right]\mathrm{sign}\left(\omega\right)
\te
For example, Kolmogorov turbulence corresponds to

\be
\mathcal{D}_K=\frac1\pi\frac{\left|\omega\right|}{\left[\omega^2+\nu\left[k\right]^2\right]}
\te
and also

\be
\gamma =\frac1\pi\:\mathrm{Im}\left[-\mathcal{G}_{ret}^{-1}\right]\mathrm{sign}\left(\omega\right)
\te
which in the case of Kolmogorov turbulence reduces to

\be
\gamma_K =\frac{\left|\omega\right|}\pi
\te
Next define the distribution function $F_1$ from

\be
\mathcal{G}^{++}=\pi\:\mathcal{D}\:F_1
\te
The transport equation is the identity \ref{fdt}, written as

\be
\mathcal{N}=\mathcal{G}^{-1}_{ret}\ast\mathcal{G}^{+,+}\ast\mathcal{G}^{-1}_{adv}
\label{fdt2}
\te
Working out the succesive compositions, we get

\be
\mathcal{N}-\left|\bar{\mathcal{D}}_{-,+}\right|^2\mathcal{G}^{+,+}=\frac {-i}2\left[\bar{\mathcal{D}}_{-,+}^*\left\{\bar{\mathcal{D}}_{-,+},\mathcal{G}^{+,+}\right\}-\bar{\mathcal{D}}_{-,+}\left\{\bar{\mathcal{D}}_{-,+}^*,\mathcal{G}^{+,+}\right\}+\left\{\bar{\mathcal{D}}_{-,+},\bar{\mathcal{D}}_{-,+}^*\right\}\mathcal{G}^{+,+}\right]
\label{transport2}
\te
Next write

\be
\bar{\mathcal{D}}_{-,+}=R-\Gamma
\te

\be
\bar{\mathcal{D}}_{-,+}^*=R+\Gamma
\te
So

\be
\mathcal{N}-\left(R^2-\Gamma^2\right)\mathcal{G}^{+,+}=-i\left[\Gamma\left\{R,\mathcal{G}^{+,+}\right\}-R
\left\{\Gamma,\mathcal{G}^{+,+}\right\}+\left\{R,\Gamma\right\}\mathcal{G}^{+,+}\right]
\label{transport4}
\te
Now

\be
\mathcal{D}=\frac1\pi\:\frac{\left(-i\right)\Gamma}{\left(R^2-\Gamma^2\right)}\mathrm{sign}\left(\omega\right)
\te

\be
\mathcal{G}^{+,+}=\frac{\left(-i\right)\Gamma}{\left(R^2-\Gamma^2\right)}F_1\mathrm{sign}\left(\omega\right)
\te
So finally, if $\omega\neq 0$

\be
A\left\{R,F_1\right\}+B\left\{\Gamma,F_1\right\}=\left[\mathcal{N}-\pi\gamma F_1\right]\mathrm{sign}\left(\omega\right)
\label{transport5}
\te
where

\be
A=\frac{\left(-\Gamma^2\right)}{\left(R^2-\Gamma^2\right)}
\te

\be
B=\frac{\left(R\Gamma\right)}{\left(R^2-\Gamma^2\right)}
\te

Eq. \ref{transport5} is quite involved because $A$, $B$, $\mathcal{N}$ and $\gamma$ are themselves functionals of $F_1$ through the self-energies. However, if we are only interested in the regression to Kolmogorov turbulence, we may approximate them by their values in a Kolmogorov-type flow. We show in Appendix A that this approximation is consistent with the so-called fourth-cumulant discard approximation \cite{MonYag71}. Then we get

\be
R=\nu_0\left(k^2\epsilon\right)^{1/3}
\te

\be
\Gamma=i\omega
\te

\be
A=\frac{\omega^2}{\left(\nu_0^2\left(k^2\epsilon\right)^{2/3}+\omega^2\right)}
\te

\be
B=\frac{i\nu_0\omega\left(k^2\epsilon\right)^{1/3}}{\left(\nu_0^2\left(k^2\epsilon\right)^{2/3}+\omega^2\right)}
\te
So (recall $\omega =k^0$) for $\omega >0$
\be
\frac{\nu_0\omega\left(k^2\epsilon\right)^{1/3}}{\left(\nu_0^2\left(k^2\epsilon\right)^{2/3}+\omega^2\right)}\left[
\frac{\partial F_1}{\partial T}+\frac23\frac{\omega}{k^2}\left(\mathbf{k}\cdot\nabla_{\mathbf{X}}\right)F_1\right]=N_0 k^{-d}-\omega F_1
\te
This equation describes the approach to Kolmogorov turbulence of an initially weakly nonhomogeneous flow.

\section{Final Remarks}

The goal of this paper has been to bring to the problem of turbulence several tools from nonequilibrium quantum field theory \cite{CalHu08} which, in our view, has not been yet exploited to their full potential. Foremost among these is the 2PI CTP EA approach and the possibility of depicting strongly coupled nonequilibrium fields by means of Kadanoff-Baym equations \cite{KadBay62}.

The reason why the method is promising is because the 2PI CTP EA allows to take full advantage of the symmetries of the theory, in this case random Galilean invariance. This allows to simplify the problem to the point where simple dimensional arguments can be used to completely specify relevant quantities, such as the velovity correlations at unequal times.

On the other hand this paper only sets the stage for the generalization of these methods to the mode demanding problems of sheared and bounded flows. We intend to continue our research in this direction.

\section*{Acknowledgments}
This work is supported by University of Buenos Aires, CONICET and ANPCyT (Argentina).

\section*{Appendix A: Higher correlations and variations of the self-energies}
The above analysis shows that the self-energies are determined by the velocity and pressure correlations in a homogeneous solution. We shall now see how higher correlations may be used to determine the variations of the self-energies with respect to the velocity correlations.

Let us begin with the derivatives of the mean fields 

\be
\frac{\delta \bar{x}^{\mu}}{\delta J_{\nu}}=icG^{\mu\nu}
\te

\be
\frac{\delta \bar{x}^{\mu}}{\delta K_{\nu\rho}}=\frac{ic}2\left\{\left\langle X^{\mu}X^{\nu}X^{\rho}\right\rangle -\bar{x}^{\mu} \left[\bar{x}^{\nu}\bar{x}^{\rho}+G^{\nu\rho}\right]\right\}\equiv\frac{ic}2\left\{ C_3^{\mu\nu\rho}+\bar{x}^{\nu}G^{\mu\rho}+\bar{x}^{\rho}G^{\mu\nu}\right\}
\te
and the propagators

\bea
\frac{\delta G^{\mu\nu}}{\delta J_{\rho}}&=&\frac{\delta }{\delta J_{\rho}}\left[\left\langle X^{\mu}X^{\nu}\right\rangle-\bar{x}^{\mu}\bar{x}^{\nu} \right]\nn
&=&ic\left\{\left\langle X^{\mu}X^{\nu}X^{\rho}\right\rangle-\bar{x}^{\rho}\left[\bar{x}^{\mu}\bar{x}^{\nu}+G^{\mu\nu}\right]-\bar{x}^{\nu}G^{\mu\rho}-\bar{x}^{\mu}G^{\nu\rho}\right\}=icC_3^{\mu\nu\rho}
\tea

\bea
\frac{\delta G^{\mu\nu}}{\delta K_{\rho\sigma}}&=&\frac{\delta }{\delta K_{\rho\sigma}}\left[\left\langle X^{\mu}X^{\nu}\right\rangle-\bar{x}^{\mu}\bar{x}^{\nu} \right]\nn
&=&\frac{ic}2\left\{\left\langle X^{\mu}X^{\nu}X^{\rho}X^{\sigma}\right\rangle-\left[\bar{x}^{\mu}\bar{x}^{\nu}+G^{\mu\nu}\right]\left[\bar{x}^{\rho}\bar{x}^{\sigma}+G^{\rho\sigma}\right]\right.\nn
&-&\left.\bar{x}^{\nu}\left[C_3^{\mu\rho\sigma}+\bar{x}^{\sigma}G^{\mu\rho}+\bar{x}^{\rho}G^{\mu\sigma}\right]-\bar{x}^{\mu}\left[C_3^{\nu\rho\sigma}+\bar{x}^{\sigma}G^{\nu\rho}+\bar{x}^{\rho}G^{\nu\sigma}\right]\right\}\nn
&=&\frac{ic}2\left\{C_4^{\mu\nu\rho\sigma}+\bar{x}^{\sigma}C_3^{\mu\nu\rho}+\bar{x}^{\rho}C_3^{\mu\nu\sigma}+G^{\mu\rho}G^{\nu\sigma}+G^{\mu\sigma}G^{\nu\rho}\right\}
\tea
We can now take the derivatives of the equations of motion. At zero sources we get

\be
\left[\mathbf{D}_{\mu\nu}+i\mathbf{N}_{\mu\nu}\right]G^{\nu\sigma}+\frac12g_{\mu\nu\rho}\left[\bar{x}^{\nu}G^{\rho\sigma}+\bar{x}^{\rho}G^{\nu\sigma}+C_3^{\nu\rho\sigma}\right]=\frac ic\delta^{\sigma}_{\mu}
\label{t5}
\te
which yields the explicit representation

\be
\Sigma_{\mu\nu}=\frac{-c}2g_{\mu\sigma\rho}C_3^{\sigma\rho\tau}G^{-1}_{\tau\nu}
\te
and

\be
\left[G^{-1}_{\mu\tau}G^{-1}_{\varphi\nu}+i\frac{\delta\Sigma_{\mu\nu}}{\delta G^{\tau\varphi}}\right]C_3^{\tau\varphi\sigma}=icg_{\mu\nu\rho}G^{\rho\sigma}
\label{t1}
\te

\be
\left[\mathbf{D}_{\mu\nu}+i\mathbf{N}_{\mu\nu}+g_{\mu\nu\rho}\bar{x}^{\rho}\right] C_3^{\nu\sigma\tau}
+\frac12g_{\mu\nu\rho}\left\{C_4^{\nu\rho\tau\sigma}+G^{\rho\tau}G^{\nu\sigma}+G^{\rho\tau}G^{\nu\sigma}\right\}=0
\te

\be
-icg_{\mu\nu\rho} C_3^{\rho\sigma\tau}+\left[G^{-1}_{\mu\varpi}G^{-1}_{\varphi\nu}+i\frac{\delta\Sigma_{\mu\nu}}{\delta G^{\varpi\varphi}}\right]C_4^{\varpi\varphi\tau\sigma}+i\frac{\delta\Sigma_{\mu\nu}}{\delta G^{\varpi\varphi}}\left[G^{\varpi\sigma}G^{\varphi\tau}+G^{\varpi\tau}G^{\varphi\sigma}\right]=0
\label{t20}
\te

\subsection{Self-energy variations}
We shall now go back to the problem of estimating the self-energy variations in isotropic homogeneous turbulence. 
We have already seen that the assumpton that $\Gamma_Q$ depends on the propagators only through the random galilean invariant component $\mathcal{P}\left[G\right]$ implies that the self energies can be written as

\be
\Sigma_{r\alpha,s\beta}\left[\mathbf{x},t_x;\mathbf{y},t_y\right]=\int\:d\mathbf{z}dt\:\tilde\Sigma_{r\alpha,s\beta}\left[\mathbf{z},t\right]\delta\left(\mathbf{x-y-z}\right)\delta\left(t_x-t\right)\delta\left(t_y-t\right)
\te
where $\tilde\Sigma$ is itself a functional of $\mathcal{P}\left[G\right]$. A second variation yields 

\be
\frac{\delta\Sigma_{r\alpha,s\beta}\left[\mathbf{x},t_x;\mathbf{y},t_y\right]}{\delta G^{u\gamma,v\delta}\left[\xi,t_{\xi};\eta,t_{\eta}\right]}=\mathcal{F}_{r\alpha,s\beta;u\gamma,v\delta}\left[\mathbf{x-y},\xi -\eta, t_x-t_{\xi}\right]\delta\left(t_x-t_y\right)\delta\left(t_{\xi}-t_{\eta}\right)
\te
where we have already assumed homogeneity in time.

Our starting point is eq. \ref{t20}, rewritten as

\be
\frac{\delta\Sigma_{\mu\nu}}{\delta G^{\lambda\psi}}=\frac c2g_{\mu\nu\rho} G^{-1}_{\lambda\sigma}G^{-1}_{\psi\tau}C_3^{\sigma\tau\rho}+\frac i2\left[G^{-1}_{\mu\varpi}G^{-1}_{\varphi\nu}+i\frac{\delta\Sigma_{\mu\nu}}{\delta G^{\varpi\varphi}}\right]C_4^{\varpi\varphi\tau\sigma}G^{-1}_{\lambda\sigma}G^{-1}_{\psi\tau}
\label{t20b}
\te
Now use eq. \ref{t1}

\be
G^{-1}_{\lambda\sigma}G^{-1}_{\psi\tau}C_3^{\sigma\tau\rho}=icg_{\lambda\psi\eta}G^{\eta\rho}
-i\frac{\delta\Sigma_{\lambda\psi}}{\delta G^{\sigma\tau}}C_3^{\sigma\tau\rho}
\label{t1bb}
\te
to get

\bea
\frac{\delta\Sigma_{\mu\nu}}{\delta G^{\lambda\psi}}&=&\frac {ic^2}2g_{\mu\nu\rho} g_{\lambda\psi\eta}G^{\rho\eta}+\frac i2G^{-1}_{\mu\varpi}G^{-1}_{\varphi\nu}C_4^{\varpi\varphi\tau\sigma}G^{-1}_{\lambda\sigma}G^{-1}_{\psi\tau}\nn
&-&\frac {ic}2g_{\mu\nu\rho} \frac{\delta\Sigma_{\lambda\psi}}{\delta G^{\sigma\tau}}C_3^{\sigma\tau\rho}
+i\frac{\delta\Sigma_{\mu\nu}}{\delta G^{\varpi\varphi}}C_4^{\varpi\varphi\tau\sigma}G^{-1}_{\lambda\sigma}G^{-1}_{\psi\tau}
\label{t20c}
\tea
Observe that the first and third terms only turn on at $\mathbf{k}=0$, while the second and third contain the fourth cumulant. Therefore, under the fourth-cumulant discard approximation we conclude that the variations of the self-energies around a Kolmogorov solution vanish, as we have assumed in the text.

\section*{Appendix B: Three point correlations in homogeneous isotropic flows}

In order to use these equations to find the variations of the self energies with respect to the propagators, we need further information about the properties of the higher correlations. In this section we shall investigate the three point correlations.

If all arguments correspond to pressure fields, then translation invariance implies

\bea
C_3^{\alpha,\beta,\gamma}\left[\mathbf{x},t_x;\mathbf{y},t_y;\mathbf{z},t_z\right]&=&\int\frac{d\mathbf{p}}
{\left(2\pi\right)^d}\frac{d\omega_p}{\left(2\pi\right)}\frac{d\mathbf{q}}{\left(2\pi\right)^d}\frac{d\omega_q}{\left(2\pi\right)}\nn
&& e^{i\left[\mathbf{p}\left(\mathbf{x}-\mathbf{z}\right)-\omega_p\left(t_x-t_z\right)\right]}\:e^{i\left[\mathbf{q}\left(\mathbf{y}-\mathbf{z}\right)-\omega_q\left(t_y-t_z\right)\right]}C_{3}^{\alpha,\beta,\gamma}\left[\mathbf{p},\omega_p;\mathbf{q},\omega_q\right]
\tea
where $C_{3}^{\alpha,\beta,\gamma}$ is a scalar with the symmetries appropiate to the coresponding correlation.

In the case where one index corresponds to a velocity field, 

\be
\nabla^{\left(x\right)}_{r}C_3^{r\alpha,\nu,\rho}\left[\mathbf{x},t_x;\mathbf{y},t_y;\mathbf{z},t_z\right]=0
\te
This immediately shows that the correlations of one velocity and two pressures must vanish

\be
C_3^{r\alpha,\beta,\gamma}\left[\mathbf{x},t_x;\mathbf{y},t_y;\mathbf{z},t_z\right]=0
\te
For two velocities and one pressure let us write

\bea
C_3^{r\alpha,s\beta,\gamma}\left[\mathbf{x},t_x;\mathbf{y},t_y;\mathbf{z},t_z\right]&=&\int\frac{d\mathbf{p}}
{\left(2\pi\right)^d}\frac{d\omega_p}{\left(2\pi\right)}\frac{d\mathbf{q}}{\left(2\pi\right)^d}\frac{d\omega_q}{\left(2\pi\right)}\nn
&&e^{i\left[\mathbf{p}\left(\mathbf{x}-\mathbf{z}\right)-\omega_p\left(t_x-t_z\right)\right]}\:e^{i\left[\mathbf{q}\left(\mathbf{y}-\mathbf{z}\right)-\omega_q\left(t_y-t_z\right)\right]}C_{3}^{r\alpha,s\beta,\gamma}\left[\mathbf{p},\omega_p;\mathbf{q},\omega_q\right]
\tea
Then asking for transversality and that each index corresponds to a vector (as oppossed to a pseudo-vector) field we find

\be
C_{3}^{r\alpha,s\beta,\gamma}\left[\mathbf{p},\mathbf{q}\right]=\Delta_{\left(p\right)}^{ru}\Delta_{\left(q\right)}^{sv}\left\{A_H^{\left(\alpha,\beta\right),\gamma}\delta^{uv}+B_H^{\left(\alpha,\beta\right),\gamma}\mathbf{q}^u\mathbf{p}^v\right\}
\te
For three velocities $C_3^{r\alpha,s\beta,t\gamma}$, we write

\bea
C_3^{r\alpha,s\beta,t\gamma}\left[\mathbf{x},t_x;\mathbf{y},t_y;\mathbf{z},t_z\right]&=&\int\frac{d\mathbf{p}}
{\left(2\pi\right)^d}\frac{d\omega_p}{\left(2\pi\right)}\frac{d\mathbf{q}}{\left(2\pi\right)^d}\frac{d\omega_q}{\left(2\pi\right)}\nn
&&e^{i\left[\mathbf{p}\left(\mathbf{x}-\mathbf{z}\right)-\omega_p\left(t_x-t_z\right)\right]}\:e^{i\left[\mathbf{q}\left(\mathbf{y}-\mathbf{z}\right)-\omega_q\left(t_y-t_z\right)\right]}C_{3}^{r\alpha,s\beta,t\gamma}\left[\mathbf{p},\omega_p;\mathbf{q},\omega_q\right]
\tea
In this case we have the further constraint

\be
\left(\mathbf{p}+\mathbf{q}\right)_tC_{3}^{r\alpha,s\beta,t\gamma}\left[\mathbf{p},\mathbf{q}\right]=0
\te
Also the definition of $\epsilon$

\be
\left.\epsilon =\nabla_{\left(x\right)s}C_3^{r+,s+,r+}\left[\mathbf{x},t;\mathbf{y},t;\mathbf{z},t\right]\right|_{\mathbf{x}=\mathbf{y}=\mathbf{z}}
\te
becomes the sum rule

\be
\epsilon=i\int\frac{d\mathbf{p}}
{\left(2\pi\right)^d}\frac{d\omega_p}{\left(2\pi\right)}\frac{d\mathbf{q}}{\left(2\pi\right)^d}\frac{d\omega_q}{\left(2\pi\right)}\:\mathbf{p}_sC_{3}^{r+,s+,r+}\left[\mathbf{p},\omega_p;\mathbf{q},\omega_q\right]
\label{sumrule}
\te

\subsection{The von Karman-Howarth equation}
Let us now consider the equation, which is a particular case of \ref{t5} when $\bar{x}^{\nu}=0$

\be
\left[\mathbf{D}_{J\nu}+i\mathbf{N}_{J\nu}\right]G^{\nu\sigma}+\frac12g_{J\nu\rho}C_3^{\nu\rho\sigma}=\frac ic\delta^{\sigma}_{J}
\label{t6}
\te
We have two choices for $J$, $J=\mathbf{x},t_x,r,\pm$, and four for $\sigma$, $\sigma=\mathbf{z},t_z,\pm$ or $\sigma=\mathbf{z},t_z,s,\pm$. The most interesting is $J=\mathbf{x},t_x,r,-$ and $\sigma=\mathbf{z},t_z,t,+$.
The only subtlety is in computing $g_{J\nu\rho}C_3^{\nu\rho\sigma}/2$. Observe that if $J=\mathbf{x},t_x,r,-$, then necessarily $\nu=\mathbf{x}',t'_x,r',+$ and $\rho=\mathbf{y},t_y,s,+$. Therefore we get

\be
\frac12\left.\left\{\nabla_{\left(x'\right)t}C_3^{r+;t+,\sigma}\left[\mathbf{x}',t_x;\mathbf{y},t_x;\mathbf{z},t_z\right]+\nabla_{\left(y\right)t}C_3^{t+;r+,\sigma}\left[\mathbf{x}',t_x;\mathbf{y},t_x;\mathbf{z},t_z\right]\right\}\right|_{\mathbf{x}'=\mathbf{y}=\mathbf{x}}
\te
Let us Fourier transform these expressions with respect to $\mathbf{x-z}$ and $t_x-t_z$. If $J=\mathbf{x},t_x,r,-$ and $\sigma=\mathbf{z},t_z,t,\gamma$ we get

\be
i\mathbf{k}_s\int\frac{d\mathbf{q}}{\left(2\pi\right)^d}\frac{d\omega_q}{\left(2\pi\right)}C_{3}^{r+,s+,t\gamma}\left[\mathbf{k-q},\omega-\omega_q;\mathbf{q},\omega_q\right]
\te
and finally 
\be
DG^{r+,t+}+iN_{rs}G^{s-,t+}+i\mathbf{k}_s\int\frac{d\mathbf{q}}{\left(2\pi\right)^d}\frac{d\omega_q}{\left(2\pi\right)}C_{3}^{r+,s+,t+}\left[\mathbf{k-q},\omega-\omega_q;\mathbf{q},\omega_q\right]=0
\te
Let us make the index dependence explicit. We assume the bare noise self correlation is transverse and delta-correlated in time

\be
\mathbf{N}_{rt}\left[\mathbf{k},\omega\right]=\Delta_{rt}N_b\left[k\right]
\te
From the symmetry of the propagators,

\be
\int\:d\omega\:\omega\:G^{r+,t+}=0
\te
In the inertial range, both the bare forcing and the viscosity are negligible. We must conclude that

\be
i\mathbf{k}_s\int\frac{d\mathbf{q}}{\left(2\pi\right)^d}\frac{d\omega}{\left(2\pi\right)}\frac{d\omega'}{\left(2\pi\right)}C_{3}^{r+,s+,t+}\left[\mathbf{k-q},\omega;\mathbf{q},\omega'\right]=0
\te
For all $\mathbf{k}$ in the inertial range. On the other hand, from the sum rule \ref{sumrule}

\be
i\int\frac{d\mathbf{k}}{\left(2\pi\right)^d}\mathbf{k}_s\int\frac{d\mathbf{q}}{\left(2\pi\right)^d}\frac{d\omega}{\left(2\pi\right)}\frac{d\omega'}{\left(2\pi\right)}C_{3}^{r+,s+,r+}\left[\mathbf{k-q},\omega;\mathbf{q},\omega'\right]=\epsilon
\te
and so we are led to

\be
i\mathbf{k}_s\int\frac{d\mathbf{q}}{\left(2\pi\right)^d}\frac{d\omega}{\left(2\pi\right)}\frac{d\omega'}{\left(2\pi\right)}C_{3}^{r+,s+,r+}\left[\mathbf{k-q},\omega;\mathbf{q},\omega'\right]=\left(2\pi\right)^d\epsilon\delta\left(\mathbf{k}
\right)
\te
From isotropy, symmetry in $\left(r,s\right)$ and transversality in $\mathbf{z}$, we must have

\be
i\int\frac{d\mathbf{q}}{\left(2\pi\right)^d}\frac{d\omega}{\left(2\pi\right)}\frac{d\omega'}{\left(2\pi\right)}C_{3}^{r+,s+,t+}\left[\mathbf{k-q},\omega;\mathbf{q},\omega'\right]=\frac A{k^2}\left\{\Delta^{rt}\mathbf{k^s}
+\Delta^{st}\mathbf{k^r}\right\}
\te
Therefore

\be
A=\frac12\left(2\pi\right)^d\epsilon\delta\left(\mathbf{k}
\right)
\te

\subsection{The Kolmogorov $4/5$ Law}

We shall use this equation to verify the Kolmogorov $4/5$ Law

\be
\left\langle \left\{\mathbf{x\cdot\left[u\left(x\right)-u\left(0\right)\right]}\right\}^3\right\rangle=-\frac45\epsilon r^4
\te
Observe that the left hand side can be written as

\bea
-6\mathbf{x}_r\mathbf{x}_s\mathbf{x}_tC_3^{r+,s+,t+}\left[\mathbf{x},t;\mathbf{x},t;\mathbf{0},t\right]&=&-6\mathbf{x}_r\mathbf{x}_s\mathbf{x}_t\int\frac{d\mathbf{k}}{\left(2\pi\right)^d}\:e^{i\mathbf{kx}}\:\int\frac{d\mathbf{q}}{\left(2\pi\right)^d}\frac{d\omega}{\left(2\pi\right)}\frac{d\omega'}{\left(2\pi\right)}C_{3}^{r+,s+,r+}\left[\mathbf{k-q},\omega;\mathbf{q},\omega'\right]\nn
&=&-6\mathbf{x}_r\mathbf{x}_s\mathbf{x}_t\int\frac{d\mathbf{k}}{\left(2\pi\right)^d}\:e^{i\mathbf{kx}}\:\frac {\left(-iA\right)}{k^2}\left\{\Delta^{rt}\mathbf{k^s}
+\Delta^{st}\mathbf{k^r}\right\}\nn
&=&12\int\frac{d\mathbf{k}}{\left(2\pi\right)^d}\:e^{i\mathbf{kx}}\:\left(iA\right)\left[r^2\frac {\mathbf{kx}}{k^2}-\frac {\left(\mathbf{kx}\right)^3}{k^4}\right]\nn
\tea
Now let us call

\be
F_1=\int\frac{d\mathbf{k}}{\left(2\pi\right)^d}\:e^{i\mathbf{kx}}\:\frac A{k^2}
\te

\be
F_2=\int\frac{d\mathbf{k}}{\left(2\pi\right)^d}\:e^{i\mathbf{kx}}\:\frac A{k^4}
\te
Then

\be
\int\frac{d\mathbf{k}}{\left(2\pi\right)^d}\:e^{i\mathbf{kx}}\:\left(iA\right)\frac {\mathbf{kx}}{k^2}=\mathbf{x}\cdot\nabla F_1=r\frac d{dr}F_1
\te

and

\be
\int\frac{d\mathbf{k}}{\left(2\pi\right)^d}\:e^{i\mathbf{kx}}\:\left(iA\right)\frac {\left(\mathbf{kx}\right)^3}{k^4}=-\left\{\left(r\frac d{dr}\right)^3-3\left(r\frac d{dr}\right)^2+2r\frac d{dr}\right\}F_2
\te

Now

\be
\nabla^2F_2=-F_1
\te
and

\be
\nabla^2F_1=-\frac{\epsilon}2
\te
so
\be
F_1=\frac{-1}{12}\epsilon r^2
\te

\be
F_2=\frac{1}{240}\epsilon r^4
\te
Which is the desired result.

\end{document}